# Long term continuous radon monitoring in a seismically active area

Antonio Piersanti, Valentina Cannelli*, Gianfranco Galli

*Istituto Nazionale di Geofisica e Vulcanologia, Rome, Italy*

**ABSTRACT**

*We present the results of a long term, continuous radon monitoring experiment started in April 2010 in a seismically active area, affected during the 2010-2013 data acquisition time window by an intense micro seismic activity and by several small seismic events. We employed both correlation and cross-correlation analyses in order to investigate possible relationship existing between the collected radon data, seismic events and meteorological parameters. Our results do not support the feasibility of a robust one-to-one association between the small magnitude earthquakes characterizing the local seismic activity and single radon measurement anomalies, but evidence significant correlation patterns between the spatio-temporal variations of seismic moment release and soil radon emanations, the latter being anyway dominantly modulated by meteorological parameters variations.*

## 1. Introduction

Soil radon emanation represents a remarkably important phenomenon for multidisciplinary Earth science studies (see Gillmore et al. [2010] and Perrier et al. [2012] for a review). The World Health Organization (WHO, http://www.who.int/en/) and International Agency for Research on Cancer (IARC, http://www.iarc.fr/) have classified radon as a class 1 carcinogenic factor and it is considered by scientific community the second cause of lung cancer after cigarette smoke. On the other hand, the same radioactive nature of the gas, causing its harmfulness for human health, makes radon an extremely efficient marker of the dynamic phenomena taking place in the interior of the Earth. In this respect, radioactive elements detectors are among the most sensitive instruments because their "quantum efficiency" in detecting and measuring ionizing radiation is much higher than any other non-radioactive element detecting instrument [Semkow et al. 1994, Abbady et al. 2004]. It is important to remember that radon, due to its short half-life, displays poor intrinsic mobility and deep origin signals can probably be observed only if advection occurs. When radon flow rate increases, radon activity increases accordingly and dilution of radon by carrier gases, such as $CO_2$ and $N_2$, may occur. As a result large domains are found that can carry radon toward surface [Etiope and Martinelli 2002, Yang et al. 2003, Etiope et al. 2005].

The existence of a link between radon variations and seismogenic processes has been investigated for decades [Toutain and Baubron 1999], but its nature and properties are still open and debated issues. In recent years, new laboratory experiments gave unambiguous evidence of the link between the rock state of stress and variations in the radon emanation properties [Tuccimei et al. 2010, Mollo et al. 2011]. By now, world-wide compilations of radon emissions anomalies, that could be associated with variations in the seismic activity and/or occurrence of a single earthquake, are available in literature (see Cicerone et al. [2009] for a review). Nevertheless, investigations based on long term continuous monitoring of radon variations are much less common [e.g. Igarashi et al. 1995, Richon et al. 2003, Inan et al. 2008, Jaishi et al. 2014]. Indeed, it is likely that radon emission dynamics are influenced by meteorological parameters variations such as those of temperature, rainfall and pressure, with typical characteristic times ranging from hours to a year [Klusman and Webster 1981, Inan et al. 2012]. Even though several authors, exploiting different methodologies, investigated the correlation between gas radon and atmospheric variables [Kraner et al. 1964, Singh et al. 1988, Cigolini et al. 2009], its complete assessment is not straightforward. Besides meteorological parameters variations, there are probably other variables which affect the radon emanation such as soil permeability and soil porosity and it's clear that such dependence can be quite complicated varying from site to site and leading therefore to site-specific behaviour of the radon ema-nations [Inan et al. 2008]. Laboratory experiments have demonstrated that significative variations in permeability and porosity during a macroscopic sample rupture are connected with radon emission variations [Zhu and Wong 1997, Mollo et al. 2011].

Moreover, the meteorological influence seems to be highly significant also in particular configurations that apparently should be unaffected by external ambient, such as deep boreholes and isolated tunnels measuring sites and/or gamma acquisition of radon decays directly occurring in the rocks crystalline matrix [Zafrir et al. 2013]. Consequently, in order to unambiguously reveal and possibly quantitatively assess the external ambient contribution, it would be remarkably important to investigate the environmental conditions, when long periods of radon data continuously acquired are available. Such an approach would help to avoid misinterpretation of apparent anomalies in the detected signal, interpreted as a significant variation from the trend not justified by contextual climatic conditions.

In order to systematically characterize the short and long term patterns of temporal evolution of soil radon emissions in a seismically active area, we have installed a real time continuous monitoring station in Pietralunga, in the Italian region Umbria, northern Italian Apennines (Figure 1). The station (hereafter PTRL) has been installed in the framework of the multidisciplinary "The Alto tiBerina near fault ObservatOry" TABOO (http://taboo.rm.ingv.it), a research infrastructure devoted to study earthquakes preparatory processes [Chiaraluce et al. 2014], that started data acquisition on April 2010.

Here we present both almost four full years of data acquisition and the results of our experiment. The collected data show a remarkably complex time dependence. As evidenced by the works cited above, it is rather evident the major role played by meteorological parameters in modulating the radon emanations, but a deeper analysis indicates a possible role played also by seismogenic processes.

## 2. Seismotectonic setting

According to the interpretation of seismic reflection profiles, the monitored area is characterized by the presence of a 60 km long extensional fault system, active in the Quaternary and dominated at depth by an east-dipping low angle normal fault named Alto Tiberina Fault (ATF) [Pialli et al. 1998, Boncio et al. 2000, Mirabella et al. 2011]). The seismicity of this area during the last 20 years has been characterized by continuous activity of low-magnitude earthquakes, mainly associated with a NW-SE-oriented structure that ranges in depth from 6 km to 10 km. The Apennine main direction (NW-SE) that marks out this seismicity is consistent with known regional fault systems [Mirabella et al. 2011], among which the ATF is considered to be the most reliable in explaining field deformation and earthquakes registered in the Umbria–Marche region [Chiaraluce et al. 2007]. Remarkably, the ATF has accumulated a long term slip rate of about 1-3 mm/year in the last 2 Myear in absence of large historical events associated with this structure. Also the possibility of high fluid pressure at 4-5 km depth further motivated our investigations. In fact, the presence of active fluid processes could be at the same time an important earthquake triggering element [Miller et al. 2004, Mulargia and Bizzarri 2015] as well as an efficient radon transportation mechanism. The existence of fluid diffusion processes is supported by the evidence that within deep boreholes drilled in this area, carbon dioxide overpressure at about 85% of lithostatic load has been encountered. In addition, the isotopic signature of a large number of local springs indicates that the whole area is interested by a remarkable flux of $CO_2$ [Chiodini et al. 2004]. Some Italian representative surveys, devoted to estimate the national distribution of radon concentration in dwellings [Bochicchio et al. 2005] and in schools of Umbria region [Sabatini et al. 2011], show how radon concentration can significantly varies even at distance of few hundred of meters and depending from the building.

Starting from April 2010 a sharp increase in the micro-seismicity of the investigated area has been registered [Marzorati et al. 2014]. The earthquakes localized from 2010 up till now have evidenced a NW-SE-oriented structure (Figure 1), whose activation occurred in different sectors and during subsequent times: just in the month of April 2010 more than 800 earthquakes were detected in the northeast part of the Pietralunga area (Figure 2a) with a $M_l$ 3.8 event occurred on April 15, 2010, which was the largest recorded during the first Pietralunga sequence [Marzorati et al. 2014]. A renewal of seismic activity occurred during April 2013, but with an epicentral distribution displaced about 10 kilometers towards west with respect to 2010 Pietralunga sequence. More than 1600 earthquakes were registered between April and May 2013, all clustered at a maximum distance of 4 km from Città di Castello, with two major events on April 20 and on May 8, both $M_l$ 3.6 (Figure 2b). A new south-east (Gubbio) cluster of more than 5000 events (of which 1200 only in the month of December 2013) occurred from August to December 2013 [De Gori et al. 2015] (Figure 2c). The three major events registered in this period are: an $M_l$ 3.7 earthquake on August 26, an $M_l$ 3.9 earthquake on December 18 and an $M_l$ 4.0 earthquake on December 22 (see Table 1).

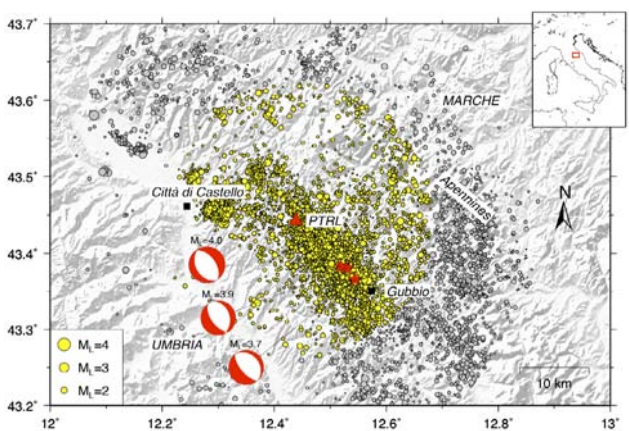

**Figure 1.** Map of the study area located in the Italian region Umbria (see inset) showing: i) the location of the radon monitoring station Pietralunga (PTRL) (red triangle); ii) the seismicity recorded by ISIDe [ISIDe 2010] between January 1, 2011 and December 31, 2013. Yellow dots represent earthquakes with epicentral distance from the PTRL station of less or equal to 20 km (8,295 events); iii) the focal mechanisms of the $M_l \geq 3.7$ earthquakes (http://cnt.rm.ingv.it/) occurred during the selected period. See Table 1 for details. The plot was made using the Generic Mapping Tools version 4.2.1 (www.soest.hawaii.edu/gmt; Wessel and Smith [1998]).

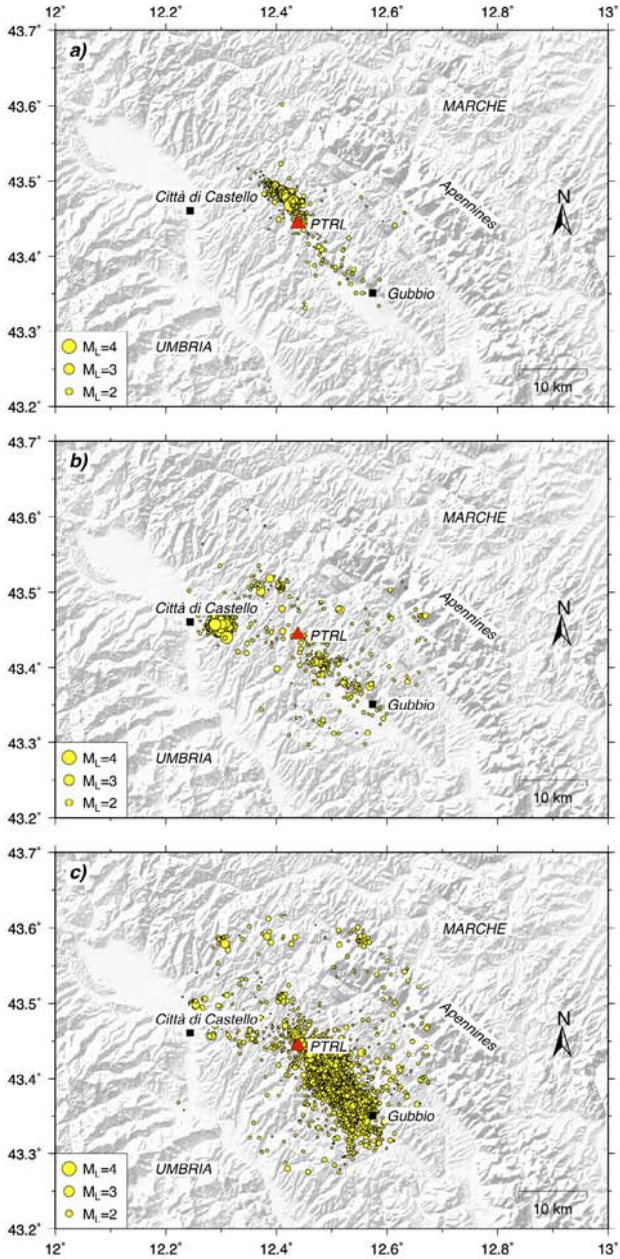

**Figure 2.** (a): Seismicity recorded by ISIDe [ISIDe 2010] between April 1, 2010 and April 30, 2010 with epicentral distance from the PTRL station (red triangle) of less or equal to 20 km (817 events); (b): the same as in panel (a) but between April 1, 2013 and May 31, 2013 (1653 events); (c): the same as in panel (a) but between August 1, 2013 and January 31, 2014 (5434 events). Plots were made using the Generic Mapping Tools version 4.2.1 (www.soest.hawaii.edu/gmt; Wessel and Smith [1998]).

Looking at the temporal evolution of the seismic moment release $M_0$ (Figure 3b), we see that the shape is fair regular with variations almost limited in a $10^2$ factor range. Incidentally, the seismicity nucleating along the ATF, including the earthquake sequences described above, is not able to explain both the short and long term deformation inferred by geological [Collettini and Holdsworth 2004] and geodetical [D'Agostino et al. 2009] studies, respectively. These observations, in absence of significant historical earthquakes, indicate that some a-seismic deformation processes are active in the area.

For the present study we have considered the seismicity satisfying the following temporal and geographical criteria: (i) occurrence time from January 1, 2011 to December 31, 2013 (transparent gray rectangles in Figures 3a and 3b); (ii) distance from PTRL station less or equal to 20 km (yellow dots in Figure 1). The resulting catalogue was obtained from the Italian Seismological Instrumental and parametric Data-basE (ISIDe) [ISIDe 2010] and consists of 8295 events, whose seismic moment release temporal evolution is shown in Figure 3b as 15-days average. The magnitude of our earthquakes dataset ranges from $M_l$ 0.1 to $M_l$ 4.0, the detectability threshold being remarkably low in this area due also to the seismic stations implemented in the framework of TABOO [Chiaraluce et al. 2014].

| EQ# | Lat | Lon | Depth | Strike | Dip | Rake | $M_l$ | Date |
|---|---|---|---|---|---|---|---|---|
| | (deg) | | (km) | (deg) | | | | (mm.dd.yyyy) |
| 1 | 43.37 | 12.54 | 8.6 | 140 | 65 | -90 | 3.7 | 08.26.2013 |
| 2 | 43.38 | 12.53 | 8.9 | 144 | 58 | -102 | 3.9 | 12.18.2013 |
| 3 | 43.38 | 12.52 | 8.3 | 324 | 52 | -94 | 4.0 | 12.22.2013 |

**Table 1.** Parameters of the $M_l \geq 3.7$ earthquakes occurring in Pietralunga area in the period between January 1, 2011 and December 31, 2013. The distributions of the earthquakes are given in Figure 1.

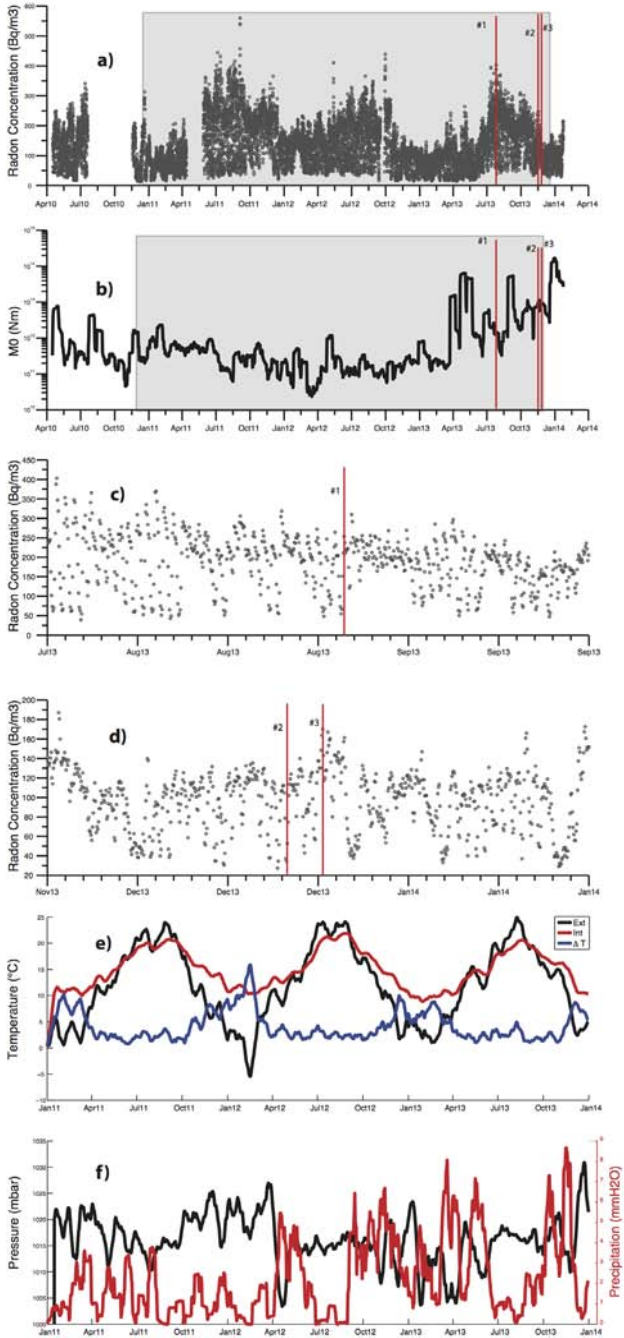

**Figure 3.** (a): Radon time serie from PTRL station between April 2010 and January 2014 (radon in counts/115 min). (b): 15-days moving averaged time serie of cumulative seismic moment release $M_0$ in the same time-window of panel (a). The transparent gray rectangles represent the time-window we selected for our analysis. Vertical red lines in panels (a) and (b) represent the occurrence of the 3 earthquakes $M_l \geq 3.7$ during the period (see Table 1) we selected for our analysis. (c): an enlarged view of panel (a) of ± 40 days around earthquake #1 (vertical red line) of Table 1. (d): the same as in panel (c) but of about ± 40 days around earthquakes #2 and #3 (vertical red lines) of Table 1. (e): 15-days moving averaged external temperature (black), internal temperature (red) and their difference (blue) between January 1, 2011 and December 31, 2013. (f): 15-days moving averaged pressure (black) and precipitation (red) between January 1, 2011 and December 31, 2013.

### 3. Radon: station and data acquisition

Soil radon emanation time serie collected by PTRL station from April 2010 up to January 2014 (with breaks due to technical failures in 2010 from 07/20 to 11/18, in 2011 from 04/12 to 05/28 and in 2012 from 09/16 to 09/28) is shown in Figure 3a. Data were acquired by a prototype station based on a Lucas cell [Lucas 1957] installed in the basement of a school (occasionally accessible only to technical staff) at Pietralunga, about 35 km north of Perugia. The site (43°26′34.7″ N, 12°26′19.5″ E) is located at about 750 m above sea level. The PTRL station has been the first installation in the framework of the TABOO research infrastructure. Up to now four prototype stations have been installed but, presently, PTRL is the only one with a radon data coverage longer than a year and including in the acquisition time window a continuous seismic activity localized in the monitored area. The radon data have been continuously acquired with a 2-h sampling time for about 45 months but, for this study, just 3 years (2011-2013) of observations are considered (transparent gray rectangle in Figure 3a), being the 2010 dataset seriously affected by acquisition failures. Given the typical sensitivity, efficiency and background noise figures for the adopted detectors, a 2-h sampling window allows for a reliable measure of radon concentrations ranging from few $10^1$ up to several $10^4$ Bq/m$^3$. The internal temperature is acquired by a specific sensor, simultaneously with radon concentration, whilst all the other meteorological parameters daily values (external temperature, pressure, precipitation) are obtained as short term (12-24 h) weather forecast by an Italian weather forecasting site (http://www.ilmeteo.it/). Also meteorological observations are presented as 15-days moving average (Figures 3e and 3f). In Appendix A we present the results obtained during an experiment performed at our test laborartory of Istituto Nazionale di Geofisica e Vulcanologia in Rome in order to test and validate our acquisition method.

### 4. Data analysis and results

In Figure 3a our whole dataset of radon observations is displayed. The measured concentrations range from less than 50 to little more than 500 Bq/m$^3$, being limited for most of the sampling intervals between 100 and 300 Bq/m$^3$. Even a coarse look at the data allows us to evidence a significant daily and seasonal component in the time series, likely associated with the typical periodical variations of the meteorological parameters. This is quantitatively immediately evident for high frequency harmonics, where a simple Fourier analysis shows a strong daily component both in temperature and radon time series (Figure 4). The lower frequency components are not such immediately evident from the Fourier analysis, even if a coarse examination of radon time serie (Figure 3a) shows similarity in the seasonal trend for 2011, 2012 and 2013: in agreement with Inan et al. [2012] the radon level appears lower during most winter-spring months (November-May) with a variability range between 100 and 200 Bq/m$^3$, and higher during summer-fall months (June-October), with slightly different variability ranges for the three years, between 300 and 550 Bq/m$^3$ for 2011, between 200 and 450 Bq/m$^3$ for 2012 and between 200 and 400 Bq/m$^3$ for 2013. Looking at the temporal trend of the internal and external temperature (Figure 3e), the major influence of the variation of these parameters with respect to the other meteorological ones on detected radon concentrations appears actually evident.

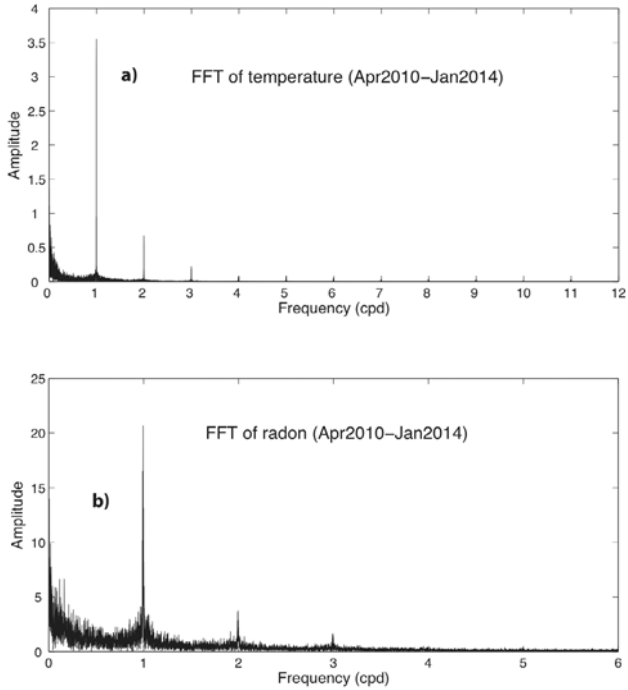

**Figure 4.** (a): FFT amplitude spectrum of the internal temperature (constant sample increment with time interval of 1 hour measured by PTRL station). (b): FFT amplitude spectrum of the radon (constant sample increment with time interval of 2 hours measured by PTRL station).

During the time window of the experiment three $M_l > 3.7$ earthquakes occurred: on August 26, 2013 ($M_l = 3.7$), December 18, 2013 ($M_l = 3.9$) and December 22, 2013 ($M_l = 4.0$) (see Table 1, Figure 1 and Figure 3b). Looking at Figures 3c and 3d, we can see that no macroscopic and common radon anomaly has been de-tected before or after those events. Nevertheless, looking at the global trend (Figure 3a), we can see that, differently from the same period of 2011 and 2012, since the end of July 2013 (approximately one month before the first $M_l$ 3.7 earthquake, Figure 3c) an overall decreasing trend of radon concentrations has been registered, starting from an absolute high level as usual in summer in our whole detection window. Focusing on the weeks before 2nd and 3rd earthquake (Figure 3d), again we see slightly peaked and then again decreasing radon concentrations about 30-40 days before the earthquake occurrence. A similar behaviour has been observed in laboratory experiments on rocks under increasing stress [Tuccimei et al. 2010, Mollo et al. 2011], where a radon release reduction is measured in compacted samples and increase of the same one is measured after failure.

In order to further investigate possible relationship between radon emanations and seismic moment release, as well as between radon emanations and the other environmental variables, we performed a wide range of correlation and cross-correlation analyses on the associated datasets. We analyzed both the unfiltered time series as well as those filtered using moving averages with different window widths (3-7-15-days). In the following, we decided to present and discuss the results for a 15-days moving average filter that, in our opinion, better highlights the relevant features, being anyhow evident with other filter techniques. All the analyses have been separately performed and presented for each year from 2011 to 2013.

### 4.1. Pearson correlation

In order to test the possible excessive rigidity of a classic correlation analysis that could reflect only the strength of a linear relationship among the investigated variables, we have performed our computations also evaluating a non parametric correlation coefficient [Kendall 1970]. Since we have verified that all our results are insensitive with respect to the considered correlation coefficient, we prefer to keep the usual Pearson coefficient [see, e.g., Snedecor and Cochran 1989].

In Figures 5, 6 and 7 the results of Pearson correlation analysis between detected radon concentrations and seismic moment release, internal, external and differential temperature, barometric pressure, and rainfall are presented for 2011, 2012 and 2013, respectively.

Each of these figures is arranged in eight subplots: all the available time series are shown in the left subplots (panels a-d), whilst the right ones represent the monthly trend of the RHO coefficient between radon concentration and the corresponding variable on the left (panels f-h), except for the subplot on the top right (panel e) that shows the correlation cumulative annual values between radon concentration and all the other variables, both as 15-days moving averages (whose numerical values are listed in the first three columns of Table 2) and residuals. In all the subplots showing correlation values, the horizontal solid lines mark the 99% significance level. In what follows we will discuss, for each couple of variables and for each temporal segments, the results of the Pearson correlation analysis.

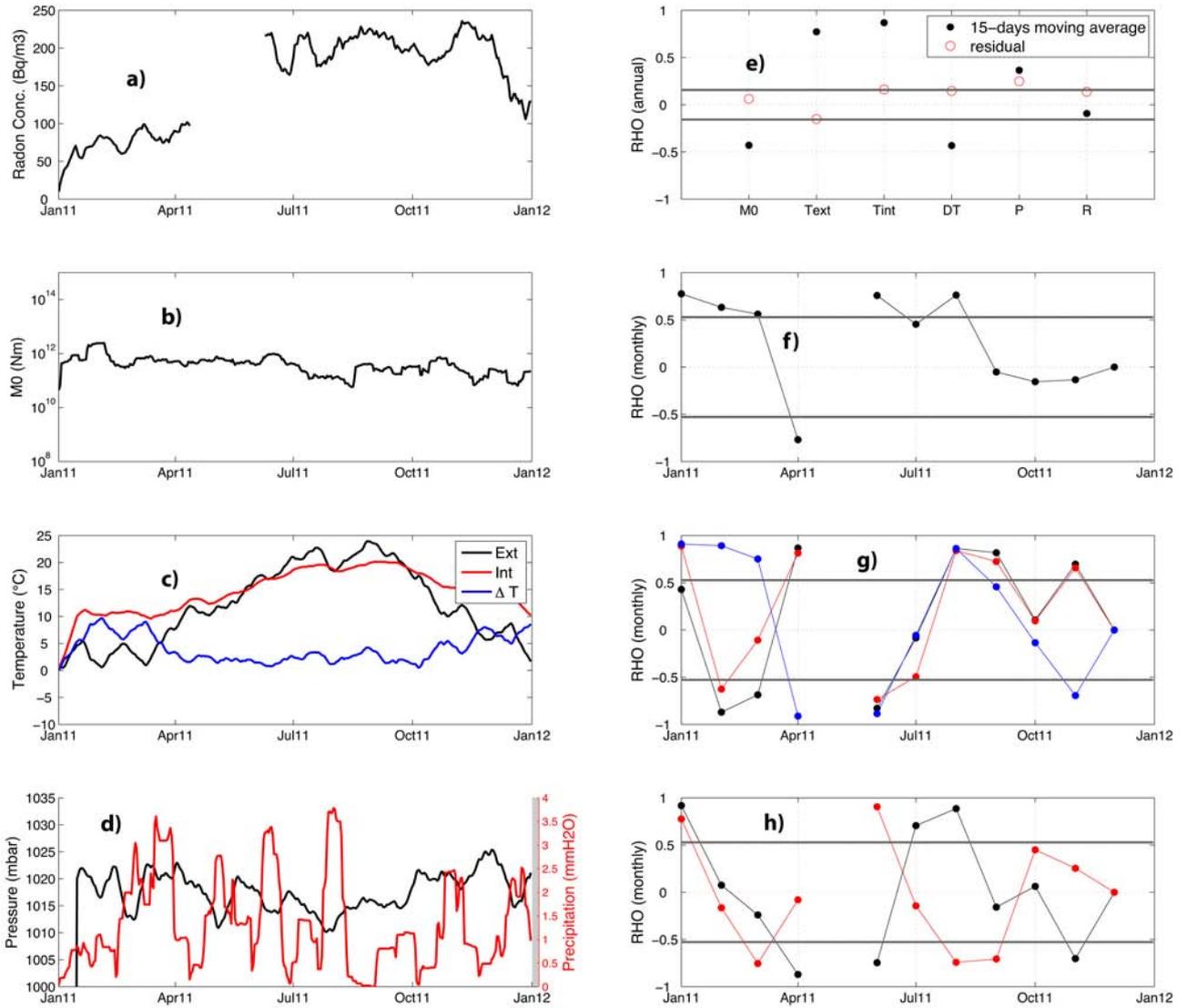

**Figure 5.** Left panels from up to bottom (a-d): 15-days moving averaged time series of radon concentration, cumulative seismic moment (in logarithmic scale), temperature (black external, red internal and blue their difference $\Delta T$), pressure/precipitation (black and red line, respectively) between January 1, 2011 and December 31, 2011. Right panels from up to bottom: (e) annual value of Pearson's correlation coefficient (RHO) between radon concentration and each one of other variables in left panels (both between 15-days moving averages and residuals obtained subtracting the 15-days moving average from the unfiltered data); (f-h) monthly trend of RHO evaluated between radon concentration and cumulative seismic moment/temperature/pressure/precipitation; horizontal gray lines represent 99% confidence threshold.

The RHO Pearson's correlation coefficient between radon and temperature confirms its role in modulating the gas emanation. Significant level of correlation are obtained both as annual and as monthly values. RHO ranges from 0.6 to 0.9 over the entire 2011, 2012 and 2013 years, being the correlation with the internal temperature always higher than correlation with the external one. The annual correlation values show that, for the three analyzed years, radon variations are positively correlated with external ($T_{ext}$) and internal ($T_{int}$) temperatures and negatively correlated with the differential one ($\Delta T$). The monthly correlation trend among radon and temperatures is more complex but for $T_{int}$ and $\Delta T$ there is an evident coherence also in the monthly correlation sign, while for $T_{ext}$ this coherence is lacking.

For what concerns radon-pressure and radon-precipitation correlation, annual values for 2011, 2012 and 2013 years are smaller than the radon temperature ones for both couples nevertheless often significant, RHO (Rn,P) always produces a positive correlation with a maximum value of 0.45 for 2012 and RHO (Rn,R) al-ways produces a negative one with a maximum value of –0.22 for 2012. This is consistent with previous investigations indicating an important but minor role of these parameters with respect to temperature in modulating radon emanations [Zafrir et al. 2013]. The monthly trends both for (Rn,P) and (Rn,R) show a large variability with no evident patterns for each considered year.

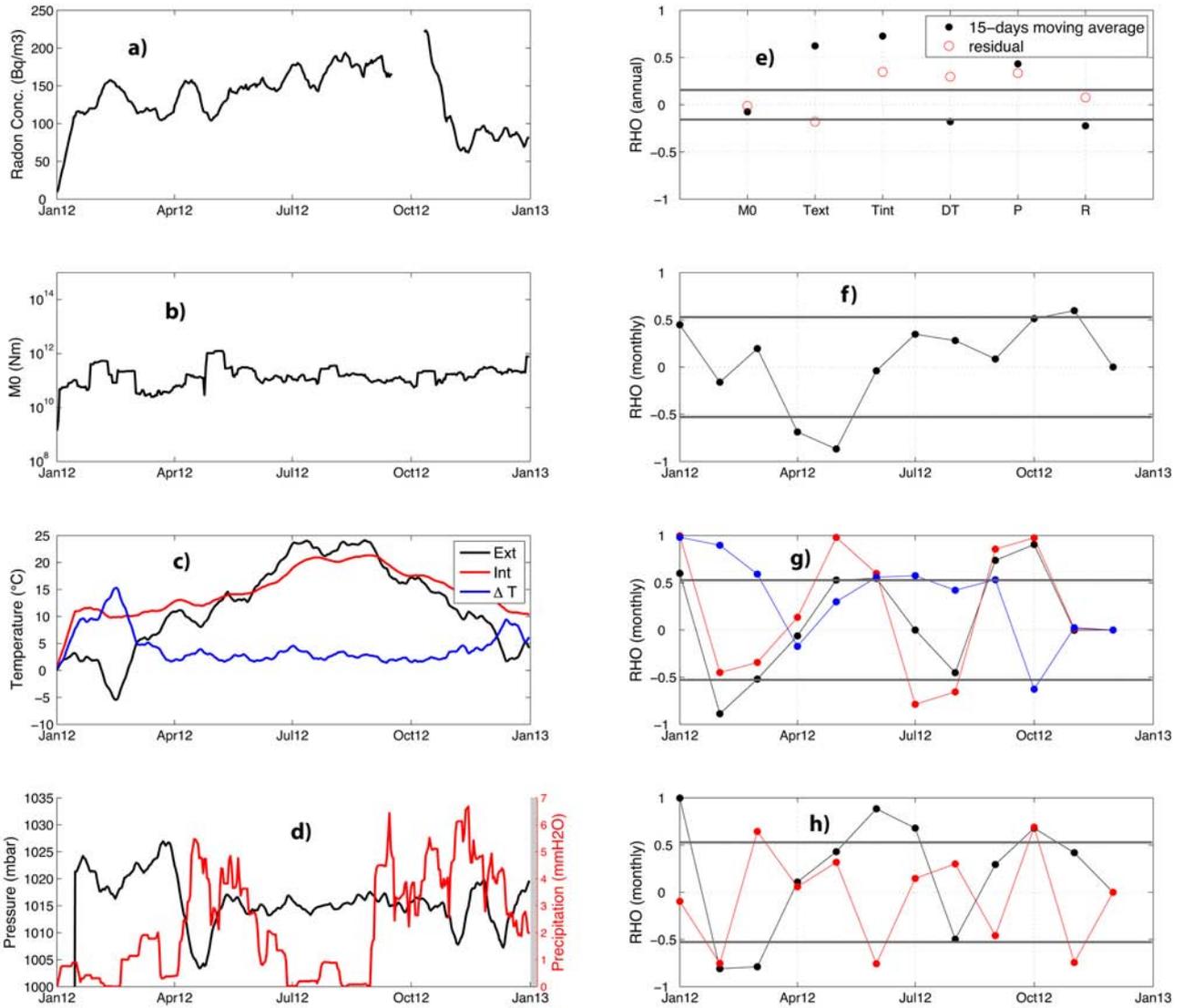

**Figure 6.** The same as in Figure 5, but between January 1, 2012 and December 31, 2012.

In Figure 4 we have evidenced the strong correlation between radon concentrations and meteorological parameters values for periods shorter than 24 h (actually we have explicitly shown it only for internal temperature for which we have hourly values). The Pearson correlation analysis for the residuals shows that for intermediate frequencies (period comprised between 1 and 15 days) the meteorological parameters variations impact is less evident in the evolution of radon observed signals.

From a qualitative analysis of the figures, we can observe that, for what concerns the global annual (Rn,$M_0$) correlation, its value (Figures 5e, 6e and 7e) is almost always not significant, except for 2011 during which the annual correlation value (–0.43) lies well above the 99% confidence interval. Looking at the monthly correlation, we see that a global, univocal pattern is not evident and that months with not significant correlation values alternate with months where RHO exceeds the 99% threshold. Also the sign of the correlation varies from month to month. Remarkably, a continuous significant trend of correlation values can be evidenced from May to September 2013 (Figure 7f), just after the reactivation of the seismic sequence in the Città di Castello area started in April 2013, though the sign of the correlation is not stable, suddenly alternating positive (May, June, September) and negative (June, July) values and always well above the 99% confidence interval. Indeed, during this time window, the cumulative seismic moment release increases by almost two orders of magnitude (Figure 3b). When the Città di

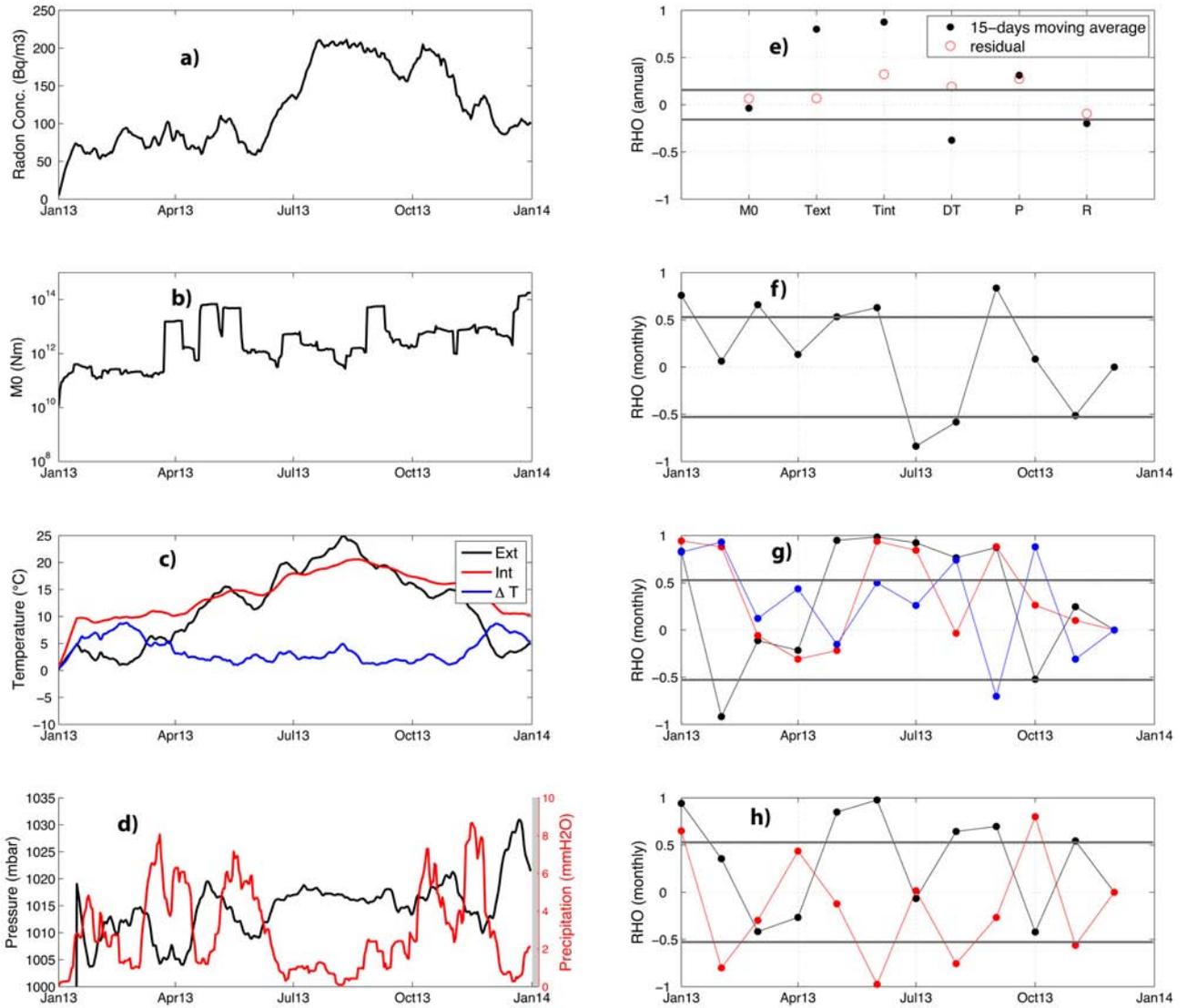

**Figure 7.** The same as in Figure 5, but between January 1, 2013 and December 31, 2013.

Castello sequence reaches its peak of activity (Figure 2b) with the two $M_l$ 3.6 events on April 20, 2013 and on May 8, 2013, the rate of seismicity suddenly increases from about 50 to 250 events per day.

*4.2. Lagged correlation*

The Pearson correlation analysis, previous described, assumes an instantaneous feedback among the investigated variables. If a causal relationship indeed exists but it is offset in time, the correlation coefficient may decrease or vanish. In such cases a simple correlation analysis can be misleading, whilst methods using the cross-correlation function are more appropriate [Box and Jenkins 1976]. In order to explore this possibility, we texted a lagged relationship between time series. Lagged correlation, that often well characterizes many natural physical systems, refers to the correlation between two time series shifted in time relative to one another. In our case, the fact that one variable may have a delayed re-sponse to the other ones is of great interest for the potential one-to-one association between seismic events and radon anomalies. We evaluated the cross-correlation function as the correlation between radon concentration and each of the other variables, shifted against one another as a function of number of observations of the offset. Formally it reads [e.g. Chatfield 2004]:

$$CC_{uy}(k) = \begin{cases} \dfrac{1}{N} \sum_{t=1}^{N-k} (u_t - \bar{u})(y_{t+k} - \bar{y}) & \star \\ \dfrac{1}{N} \sum_{t=1-k}^{N} (u_t - \bar{u})(y_{t+k} - \bar{y}) & \star\star \end{cases} \quad (1)$$

$\star \; k = 0, 1, \ldots, (N-1)$
$\star\star \; k = -1, -2, \ldots -(N-1)$

where $N$ is the series length, $u_t$ and $y_t$ are the two time series, $\bar{u}$ and $\bar{y}$ are their sample means, and $k$ is the lag. Differently from Pearson linear correlation, the cross-correlation coefficient is not normalized *a-priori*: in order to grant compatibility with the previous analyses, we normalized the cross-correlation coefficient here so that it varies between –1 and 1 and set the lag range between –40 and 40 days.

|  | 2011 | 2012 | 2013 | 2011 | 2012 | 2013 |
|---|---|---|---|---|---|---|
|  |  | RHO |  |  | max CC @ lag |  |
| $(Rn, M_0)$ | -0.430 | -0.075 | -0.036 | -0.63 @ -22 | -0.23 @ -33 | -0.10 @ 24 |
| $(Rn, T_{ext})$ | 0.772 | 0.627 | 0.799 | 0.77 @ 0 | 0.62 @ 0 | 0.80 @ 0 |
| $(Rn, T_{int})$ | 0.871 | 0.735 | 0.878 | 0.87 @ 0 | 0.73 @ 0 | 0.88 @ 0 |
| $(Rn, \Delta T)$ | -0.427 | -0.178 | -0.374 | -0.56 @ 22 | -0.50 @ -36 | -0.44 @ -14 |
| $(Rn, P)$ | 0.371 | 0.446 | 0.318 | 0.36 @ 0 | 0.43 @ 0 | 0.31 @ 0 |
| $(Rn, R)$ | -0.102 | -0.217 | -0.194 | -0.32 @ -13 | -0.54 @ 26 | -0.44 @ 22 |

**Table 2.** Annual value of Pearson's correlation coefficient (RHO) and maximum value of Cross-Correlation (max CC) with its respective lag between the 15-days moving average of radon concentration (Rn) and the 15-days moving average of each one of other variables (cumulative seismic moment release $M_0$, external temperature $T_{ext}$, internal temperature $T_{int}$, their difference $\Delta T$, pressure P and precipitation R).

In Equation (1) $u_t$ represents the radon time serie, whilst $y_t$ represents one of other variables time series among cumulative seismic moment release $M_0$, temperature (internal, external and their difference $\Delta T$), pressure and precipitation. So, according to this formalism, the maximum (or minimum if the time series are negatively correlated) of the cross-correlation function indicates the point in time where the time series are best aligned and in our case the corresponding $k$ (lag) represents the time delay between radon and one of the other variables time series, forward if $k > 0$ and backwards if $k < 0$.

Figures 8, 9 and 10 show the results we found for cross-correlation analysis between detected radon concentrations and seismic moment release, internal and external temperature, barometric pressure, and rainfall for 2011, 2012 and 2013, respectively. Each of these figures is arranged in six subplots and each subplot represents the cross-correlation function (CC) between radon concentration and all the other variables both as unfiltered data, 15-days moving average (whose maximum values with their respective lags are listed in the last three columns of Table 2) and their residuals.

The CC(Rn,$M_0$) correlation function shows quite a different behaviour during 2011, 2012 and 2013: only in 2011 significant negative cross-correlation extends over several lags. As for the Pearson correlation analysis, also the CC maximum values are always negative with values ranging from –0.10 to –0.63. Except for 2011, these values are only marginally significant but remarkably the correlations tend to increase with a negative lag for all the three years indicating that some radon signal could anticipate the moment release variations (see also Figures 3c and 3d).

The cross-correlation function between radon and temperature, both external and internal, indicates clearly that these two variables are significantly positively correlated, exceeding the 99% confidence threshold over all the considered range and for each year. The values peaked at lag = 0 confirm that the radon emanation reaction to the meteorological variations are almost instantaneous. Also the CC(Rn,P) correlation function shows, for all three years, a positively correlated trend, even if with much lower values, that not always exceed the 99% confidence threshold, except for the pick value that ranges from 0.3 in 2013 to 0.4 in 2012 and always at a corresponding lag = 0.

For what concerns the CC(Rn,R) correlation function, the results show a similar trend only for 2012 and 2013, for which the lags corresponding to their cross-correlation maximum values are comparable: lag = 26 for 2012 and lag = 22 for 2013. As for the Pearson correlation, analysis also the lagged correlation always produces a negative maximum values ranging from –0.3 in 2011 to –0.5 in 2012.

## 5. Conclusive remarks

We presented the results of a long term, continuous radon monitoring experiment performed in a seismically active area, interested by an intense micro seismic activity and by several small seismic events occurred during the experiment time window (2010-2013). The time span of our database allowed us to highlight the different roles played by the relevant variables in modulating the radon emanation levels involving time scales ranging from hours to a year. In fact, it has been shown that, by monitoring a site for at least one year, the site-characteristic patterns of soil gas radon emanation behaviour have become evident. In this respect, Fourier and Pearson correlation analysis clearly shows that meteorological parameters play a prominent role with respect to seismic moment release. It's also confirmed the known dominant role of the temperature in modulating the gas emanation. Nevertheless, correlation analysis indicates that in periods of (relatively) major seismic activity, also internal seismogenic processes could play a significant role.

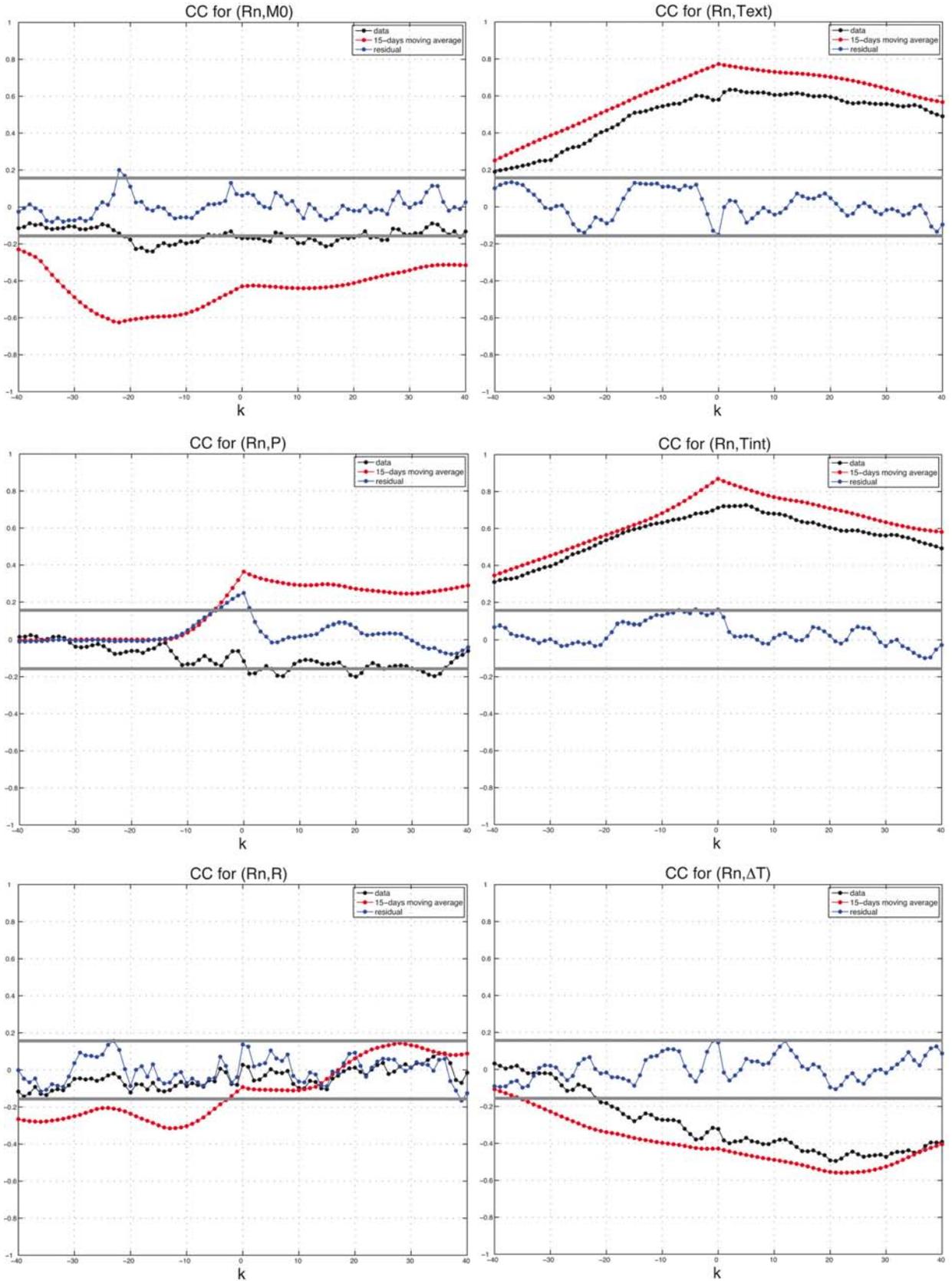

**Figure 8.** Cross-correlation function (CC) for the segment January 1, 2011 - December 31, 2011 between radon time series Rn and time series of one of other variables (cumulative seismic moment release $M_0$, temperature ($T_{ext}$, $T_{int}$ and their difference $\Delta T$), pressure P and precipitation R). The CC is evaluated between unfiltered data (black-pointed line), 15-days moving average (red-pointed line) and their residuals (blue-pointed line). Horizontal gray lines represent 99% confidence threshold.

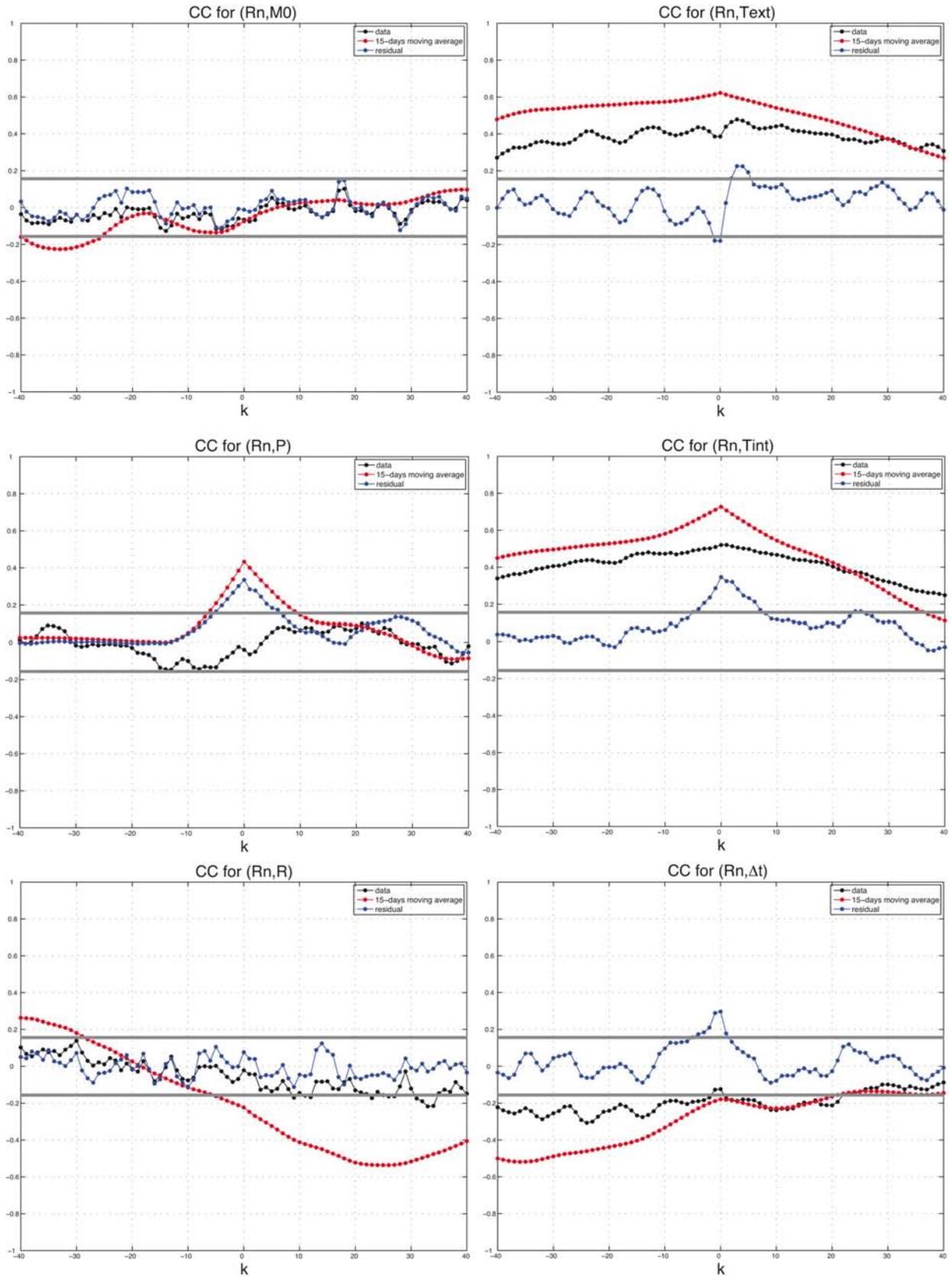

**Figure 9.** The same as in Figure 8, but for the segment January 1, 2012 - December 31, 2012.

Though during the time-window of our experiment no significant seismic event occurred, for what concerns the major seismic events registered in this period (maximum magnitude $M_l$ 4.0), our analyses did not evidence a quantitatively robust and univocal one to one association between a single event and a specific radon observation anomaly. Nevertheless, it is interesting to note a slight peak and then decay of the measured radon levels approximately 30-40 days before the occurrence of the three major seismic events of the experiment time window, especially considering the expected soil radon background behaviour of seasonal increase in early summer months and the following fall from summer to winter. Remarkably, the decreasing trend is similar to that detected in laboratory experiments before failure and the 30-40 days time-shift is similar to that maximizing the cross-correlation function between radon and seismic moment release time series.

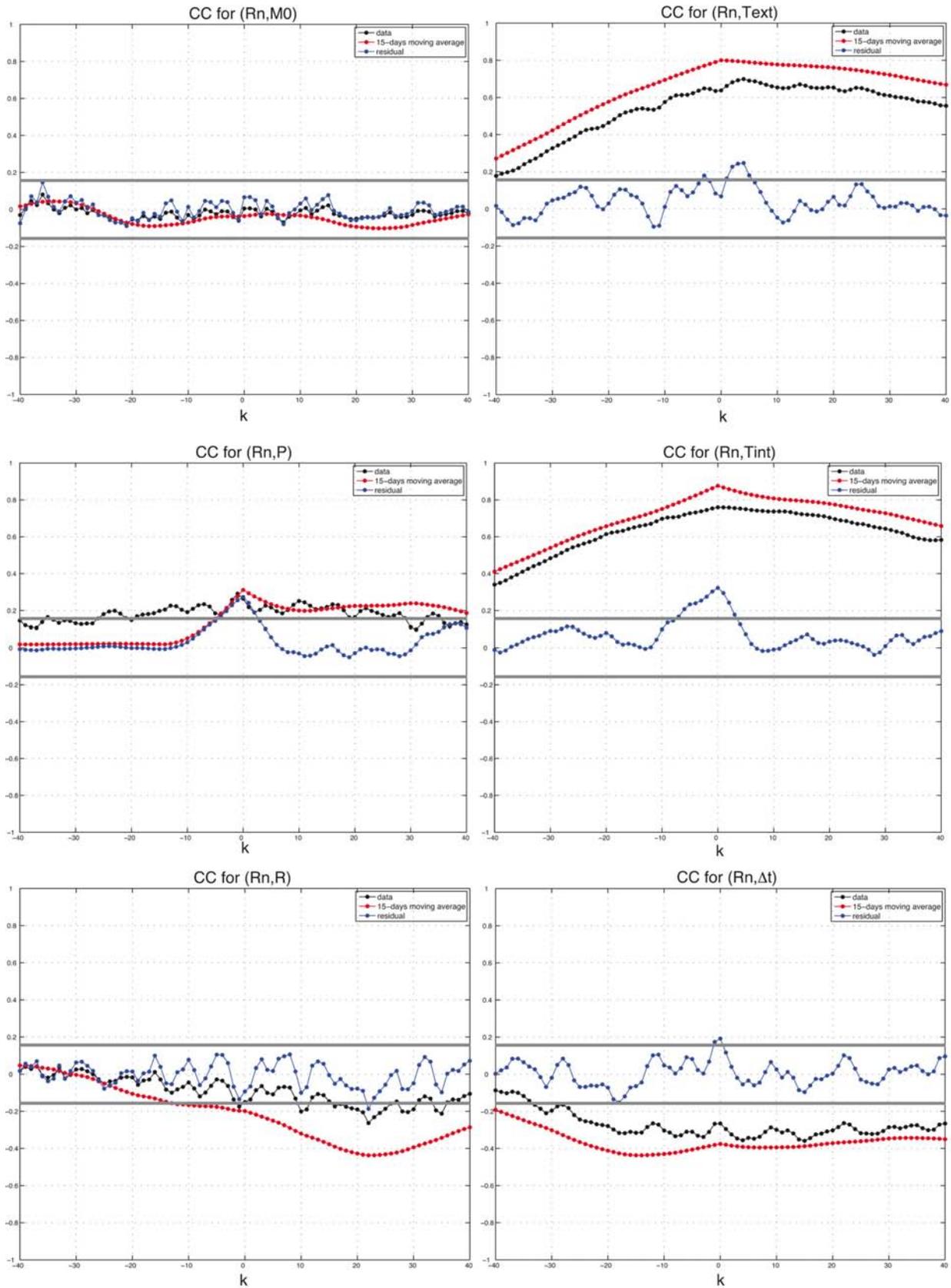

**Figure 10.** The same as in Figure 8, but for the segment January 1, 2013 - December 31, 2013.

Our results are still far from giving conclusive answers but clearly indicate that the availability of long term continuous time series of all the relevant variables becomes a crucial requirement in order to seek a robust assessment of the link between seismogenic processes and radon emissions anomalies. This is likely due to the very physics of the internal seismogenic processes that involves many different time scales, from months to centuries, but also to the raw quantitative statistical significance of a correlation or cross-correlation coefficient that is function of not only the size of correlation but also of the length of the analyzed period. For the same level of correlation, the longer the time series, the greater the statistical significance.

Our observational setup is presently unable to look for a correlation between radon and carrier gases, such as $CO_2$ and $N_2$. Anyway, when radon concentrations are low (as is the case of PTRL), even state of the art instruments have not enough sensitivity to measure carrier gases variations. On the contrary in big spaces, as caves [Batiot-Guilhe et al. 2007, Kowalczk and Froelich 2010] and tunnels [Perrier et al. 2007, Perrier et al. 2009, Perrier and Richon 2010], where radon concentrations are very high, it is possible to monitor also carrier gases. Moreover our PTRL station, investigated radon concentration temporal evolution, isn't *a-priori* influenced by permeability and porosity of the soil, while it's undoubted that these soil characteristics become necessary in case of areal measurements campaigns, especially if the radon detector has a probe inserted directly in the soil. At the same time, the availability of such "rich" data-sets requires the development of suitable post-processing and computational procedures, in order to fully exploit their informative potentialities.

Our next future development will be to implement a network of permanent radon measuring stations in the ATF area, in order to investigate the mutual relationships among the different registered signals and possibly to exploit them in order to improve our knowledge over the seismogenic processes and earthquake physics.

**Acknowledgements.** We thank Francesca Bianco, the A.E. and two anonymous reviewers for very constructive suggestions which have helped to improve the manuscript. This work was partly supported by MIUR (Ministero dell'Istruzione, dell'Università e della Ricerca) with the Premiali 2011 grant "Studio multidisciplinare della fase di preparazione di grandi terremoti".

**Data and sharing resources**

The earthquake catalogue used in this study was obtained from: http://iside.rm.ingv.it/. The meteorological parameters daily values are obtained from: http://www.ilmeteo.it/.

*Corresponding author:* Valentina Cannelli,
Istituto Nazionale di Geofisica e Vulcanologia, Rome, Italy;
email: valentina.cannelli@ingv.it.

PIERSANTI ET AL.**Appendix A. Measurements set up**

The radon detector of PTRL station is located in a small room of a school basement, not disturbed by anthropogenic influences and without any kind of opening and/or aeration system. In a more classical geochemical approach, gas emanations are measured inserting a probe directly in the soil. The latter approach is probably more efficient in maximizing the magnitude of the observed signals, this being an important requirement when the sensibility of the detection system is a critical factor (this is often the case with geochemical campaigns aimed to measure several different gases). The high sensitivity of the Lucas cell installed in PTRL (0.07 cpm /(Bq m$^{-3}$)) allows us to measure radon concentrations as low as few tens Bq/m$^3$ with negligible errors. On the other hand, in the present investigation, we are interested in resolving small temporal changes of radon signal affected by seismotectonic/seismogenic processes, likely taking place some kilometers away (both in vertical and horizontal dimensions) from the observation site and not in the local absolute radon concentrations. In this respect, the relative differences between a measurement equipment for monitoring soil radon activity with probe in air with respect to another with probe in soil does not play a crucial role.

In Figure 1A we show the radon signals simultaneously measured during an experiment performed at our test laboratory of INGV in Rome, where we installed a couple of twin stations using both techniques: a detector located in a small room in the basement (INGI) and an identical detector with a probe inserted directly in the soil (INGO), less than 1 m away. The plotted blue and red curves display 15-days moving averages from INGI and INGO stations, respectively, that is the filter used in all our statistical computations. Apart from the difference in the absolute magnitude, the two signals show remarkably similar trend and relative variations, that is to say a transfer functions exists that translates one signal into the other. Incidentally, we remind that the explicit expression of such a function has no impact at all in a statistical correlation and cross-correlation analysis. Another interesting characteristic that Figure 1A seems to reveal is that the indoor signal (red line) shows apparently a greater dynamic range (i.e. relatively lower minima and higher maxima). This interesting feature could be related to the filtering effect exerted by the concrete floor with respect to the very local radon emanations. All the considerations above are confirmed by Figure 2A, showing the quantitative relationship (a scatter plot) between simultaneous measurements acquired by the twin stations. It is evident the existence of a well defined functional relation between the two different measures and also the lower dynamic range exhibited by the probe approach (i.e. the decrease in the curve slope).

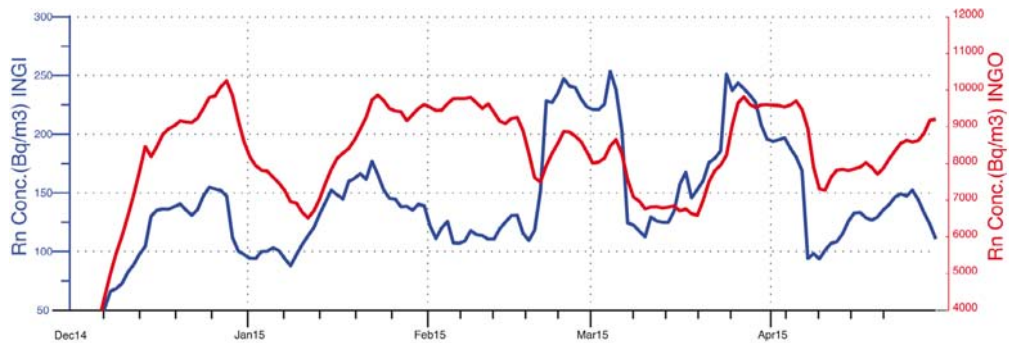

**Figure 1A.** 15-days moving averaged time series of the radon concentration from INGI (blue line) and INGO (red line) stations during the period from January 2015 to April 2015.

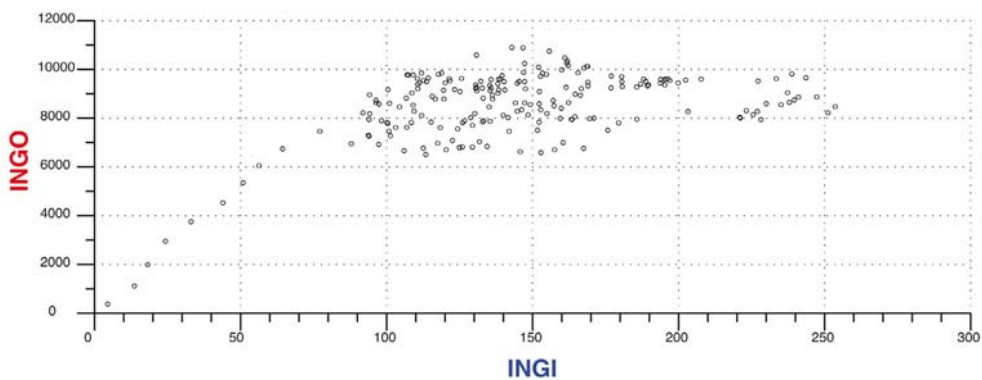

**Figure 2A.** INGO radon concentration versus INGI radon concentration during the period from January 2015 to April 2015.